\documentclass[a4paper, amsfonts, amssymb, amsmath, reprint, nofootinbib]{revtex4-2}
\usepackage[english]{babel}
\usepackage[utf8]{inputenc}
\usepackage[colorinlistoftodos, color=green!40, prependcaption]{todonotes}
\usepackage[pdftex, pdftitle={Article}, pdfauthor={Author}]{hyperref} 
\usepackage{caption}
\usepackage{subcaption}
\usepackage{amsthm}
\usepackage{mathtools}
\usepackage{physics}
\usepackage{xcolor}
\usepackage{graphicx}
\usepackage[left=23mm,right=13mm,top=35mm,columnsep=15pt]{geometry} 
\usepackage{adjustbox}
\usepackage{placeins}
\usepackage[T1]{fontenc}
\usepackage{lipsum}
\usepackage{csquotes}

\begin{document}
\title{Comparing Two-Qubit and Multi-Qubit Gates within the Toric Code}

\author{David Schwerdt}
    \email{david.schwerdt@weizmann.ac.il} 
\author{Yotam Shapira}
\author{Tom Manovitz}
\author{Roee Ozeri}
    \affiliation{Department of Physics of Complex Systems and AMOS, Weizmann Institute of Science, Rehovot 7610001, Israel}

\begin{abstract}
In some quantum computing (QC) architectures, entanglement of an arbitrary number of qubits can be generated in a single operation. This property has many potential applications, and may specifically be useful for quantum error correction (QEC). Stabilizer measurements can then be implemented using a single multi-qubit gate instead of several two-qubit gates, thus reducing circuit depth. In this study, the toric code is used as a benchmark to compare the performance of two-qubit and five-qubit gates within parity-check circuits. We consider trapped ion qubits that are controlled via Raman transitions, where the primary source of error is assumed to be spontaneous photon scattering. We show that a five-qubit Mølmer-Sørensen gate offers an approximately $40\%$ improvement over two-qubit gates in terms of the fault tolerance threshold. This result indicates an advantage of using multi-qubit gates in the context of QEC.
\end{abstract}

\maketitle

\section{Introduction} \label{sec:Introduction}
Any long-term realization of a quantum computer will need to incorporate quantum error correction (QEC). Without QEC, the performance of any given quantum algorithm will be limited by the physical qubit error rate. While much experimental effort is devoted to lowering this rate \cite{fid1,fid2,fid3}, the current values are too high to solve any large-scale problems with a direct approach. In a QEC code, information is not stored directly in physical qubits, but rather in logical qubits. The performance of the code is dictated by the fault tolerance (FT) threshold; as long as physical error rates can be kept below this threshold, then logical errors can be arbitrarily suppressed by increasing the size of the code \cite{Lidar}. \par
From an experimental standpoint then, the best QEC codes are those that have the highest FT threshold. Topological QEC (TQEC) codes \cite{K1,K2}, which encode logical qubits in global degrees of freedom, have proven to have high FT thresholds. One popular example of TQEC is the toric code, a stabilizer code \cite{Gottesman} where the structure of interactions between physical qubits maps onto the surface of a torus \cite{K3}.  Depending on the error model and simulation approach used, the threshold for the toric code (and similarly for its planar generalization surface code) has been placed between $10^{-3}$ - $10^{-2}$ \cite{Wang,Fowler,Fowler2}. \par
An accurate calculation of the threshold requires careful consideration of the physical qubit architecture. To that end, studies of QEC in a trapped ion system have been done \cite{Innsbruck,Ionqec}. Recent work considers Zeeman and hyperfine qubits and compares their performance in the toric code for various levels of magnetic field noise \cite{Brown1,Brown2,Brown3}. Other work determines how various error models affect the surface code FT threshold \cite{Stephens}.
Generally these works have considered a gate-set involving two-qubit entangling operations. However, in a trapped ion system it is possible to entangle an arbitrary number of qubits in a single gate. In the toric code, the total number of required entangling operations can be significantly reduced by using five-qubit rather than two-qubit gates. On the other hand, such multi-qubit gates are generally more susceptible to noise and error propagation. Because of this trade-off, it is hard to predict which method yields better results. We are therefore interested in the question of how using five-qubit gates influences logical error rates and the FT threshold. \par

While QEC circuits involving five-qubit gates have been suggested \cite{Innsbruck}, the implications on the toric code FT threshold have not been explored. Here we study properties of two-qubit and five-qubit Mølmer-Sørensen (MS) gates, including error propagation. We make several assumptions about the experimental setup to form an error model, and simulate the toric code for two different gate-sets: one involving two-qubit gates and one involving five-qubit gates. We extract the FT threshold in both cases, and compare their respective performance.
Here we show that the five-qubit gate method gives an improvement of approximately $40\%$ in the FT threshold, as well as a $3.7$ times improvement in logical error rates at low physical error probabilities.

\section{The Toric Code} \label{sec:The Toric Code}
\begin{figure}
\begin{subfigure}[h]{\columnwidth}
    \centering
   \includegraphics[width=.78\linewidth]{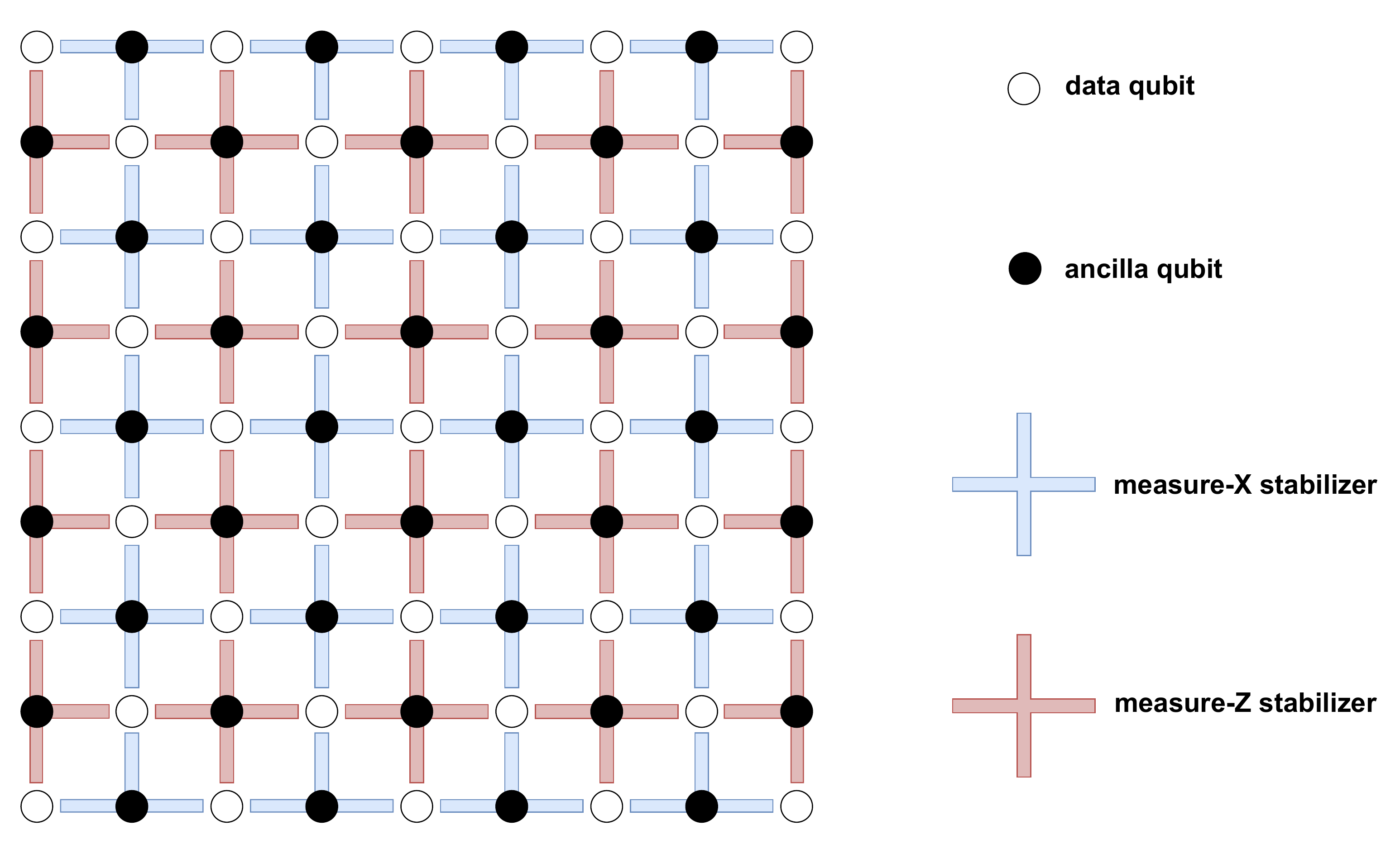}
   \caption{}
   \label{fig:sc} 
\end{subfigure}

\begin{subfigure}[h]{\columnwidth}
   \centering
   \includegraphics[width=.78\linewidth]{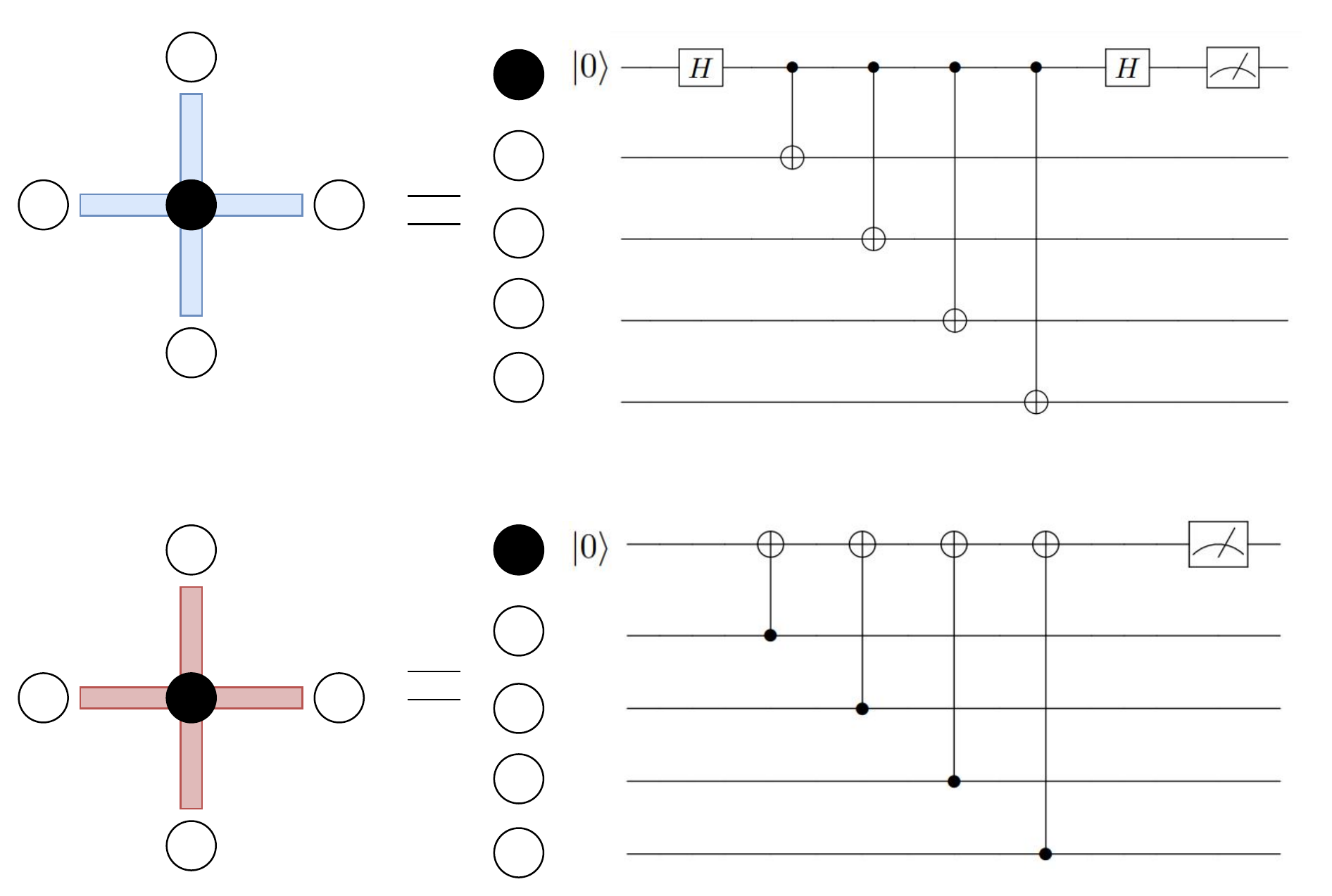}
   \caption{}
   \label{fig:sccircuit}
\end{subfigure}
\captionsetup{justification=raggedright,singlelinecheck=false}
\caption{(a) Surface code layout - white circles represent data qubits and black circles ancilla qubits. (b) Stabilizer measurement circuits - H is a Hadamard gate. At the beginning of the circuit the ancilla is initialized to the 0-state, and at the end of the circuit, it is measured.}
\end{figure}

In TQEC qubits are arranged (schematically, not necessarily physically) on a lattice, where local stabilizer operators are continually measured to infer the location of errors. The codes are characterized by the number of logical qubits they encode as well as their distance. A toric code of distance $d$ encodes two logical qubits; it is composed of a $d\cross d$ square lattice of data qubits and a $d\cross d$ square sublattice of ancilla qubits - as pictured in Figure \ref{fig:sc}. It has periodic boundary conditions in both directions so that qubits on opposing edges are identified.

Stabilizer measurements, or parity-checks, are associated with each ancilla and its surrounding four data qubits. A schematic circuit diagram for measure-$X$ and measure-$Z$ stabilizers, following the standard composition involving two-qubit CNOT gates, is shown in Figure \ref{fig:sccircuit}. \par

The stabilizer-measurement results, attained from readout on the ancilla qubits, are altered by any single $\{X,Y,Z\}$ error on data qubits. The corresponding error syndrome is decoded using a classical algorithm, for example minimum weight perfect matching (MWPM), to determine where to apply correction operators. A logical $X$ ($Z$) error occurs when a sequence of physical $X$ ($Z$) errors and their resulting correction operations form a closed, irreducible, chain along the torus. \par

\section{Stabilizer Measurement with Multi-Qubit Gates} \label{sec:Stabilizer Measurement with Multi-Qubit Gates}
Stabilizer measurement circuits are generally composed of two-qubit CNOT gates between the ancilla and its adjacent data qubits. For qubit architectures that permit only local two-qubit interactions, this is the only reasonable implementation. In the case of trapped ions, however, global interactions between any set of qubits within the ion register are possible (regardless of physical proximity). This is because entanglement is mediated by the ions’ common motional modes, as in the Mølmer-Sørensen (MS) interaction \cite{MS1,MS2}. \par
A two-qubit MS gate between ions $i$ and $j$, with entangling phase $\theta_{ij}$, results in the unitary operation,
\begin{equation}
    XX_2(\theta_{ij}) = e^{-i\frac{\theta_{ij}}{2}X_iX_j}.
    \label{eq:MS2}
\end{equation}
While a multi-qubit ($n > 2$) MS gate is given by,
\begin{equation}
    XX_n(\vec{\theta}) = e^{-i\sum _{i \neq j}\frac{\theta_{ij}}{2}X_iX_j}.
    \label{eq:MS5}
\end{equation}

For two qubits, the MS gate is equivalent to CNOT up to single-qubit rotations \cite{Maslov1}.

\begin{figure}
\centering
\begin{subfigure}[h]{\columnwidth}
   \includegraphics[width=.78\linewidth]{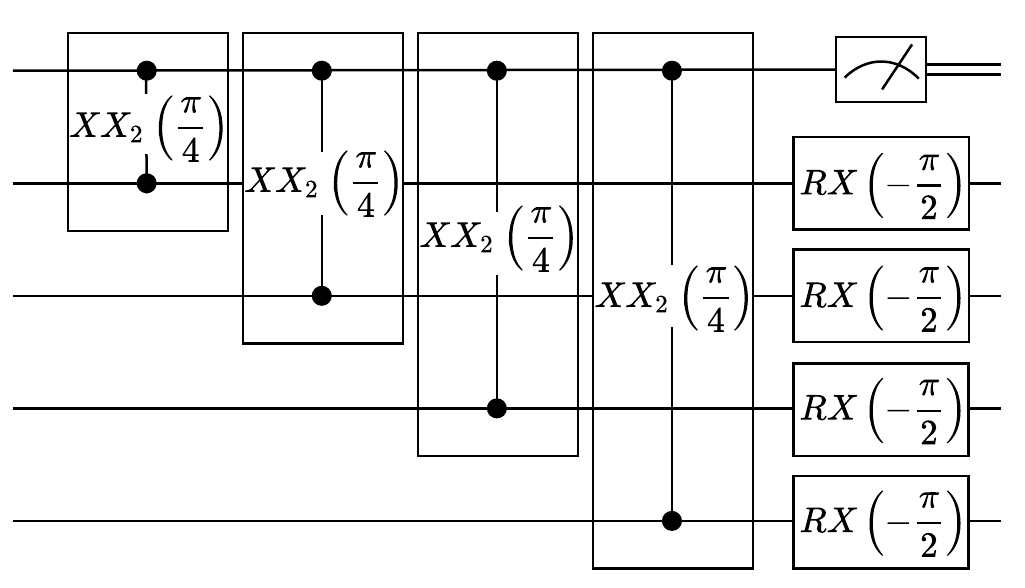}
   \caption{}
   \label{fig:4ms} 
\end{subfigure}

\begin{subfigure}[h]{\columnwidth}
   \includegraphics[width=.78\linewidth]{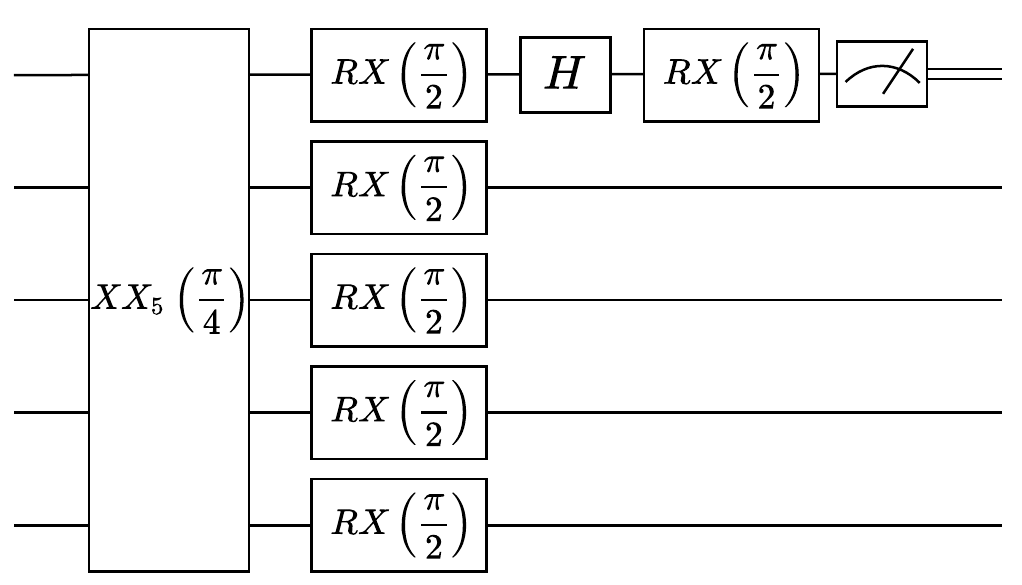}
   \caption{}
   \label{fig:1gms}
\end{subfigure}
\captionsetup{justification=raggedright,singlelinecheck=false}
\caption[GMS]{(a) Measure-X stabilizer circtuit built from four two-qubit MS gates. The ancilla is initialized to the
0-state before this circuit begins. (b) Measure-$X$ stabilizer circuit using only one five-qubit MS gate.}
\end{figure}

In implementing the toric code, it is possible to perform gates involving all five ions involved in a given stabilizer measurement. We would like to determine how the toric code performance changes when using either two-qubit or five-qubit MS gates within stabilizer measurement circuits. A comparison of measure-$X$ stabilizer circuits using two-qubit and five-qubit MS gates can be seen in Figures  \ref{fig:4ms} and \ref{fig:1gms}. 

The first circuit uses two-qubit MS gates; it is an extension of the CNOT-based circuit from Figure \ref{fig:sccircuit}. Here $RX$ is a single-qubit rotation about the x-axis. The second circuit contains a single five-qubit MS gate. While this gate produces all-to-all coupling between all five qubits, many of these couplings cancel \cite{LW}. A detailed explanation of how this circuit amounts to a stabilizer measurement can found in Appendix A. A measure-$Z$ circuit is implemented similarly - the only difference being the addition of Hadamard gates on all data qubits at the beginning and end of the circuit. \par
One obvious benefit of using a five-qubit gate over the conventional two-qubit gate method is reduction in circuit depth, the total number of sequential gates required to perform the stabilizer measurement. Consequently a shorter overall circuit time can reduce the probability of error. On the other hand, making use of a multi-qubit operation inherently involves the possibility of multiple propagated errors. This is discussed further in section \ref{sec:Master Equation Simulation for Multi-Qubit MS Gates}. It should be stressed that five-qubit gates are chosen due to the five qubits of a toric code stabilizer measurement. This analysis can, however, extend to other QEC codes, where the size of the multi-qubit gate should reflect the structure of the code.\par
Before proposing an error model and attempting to quantify this comparison, it is worth specifying a standard experimental procedure for implementing multi-qubit MS gates.

\section{Physical Implementation of Multi-Qubit Gates} \label{sec:Physical Implementation of Multi-Qubit Gates}
In a typical trapped ion setup, using for example a linear RF Paul trap, ions are oriented in a long 1D chain - where specific ions can be targeted by means of individual laser addressing. A MS gate is implemented using a bichromatic laser field that drives a force dependent on the spin of the illuminated ions. The set of ions involved in a given multi-qubit MS gate can thus be controlled by turning on individual addressing beams at the desired locations. \par
The MS interaction Hamiltonian can be written (after applying a rotating wave approximation) as \cite{thesis}: \par
\begin{equation}
    H(t) = \hbar \eta \Omega (ae^{i\delta t} + a^\dagger e^{-i\delta t})\cdot \left(\sum_{i=1}^n X_i\right),
    \label{eq:MSH}
\end{equation}
where $\eta$  is the Lamb-Dicke parameter, $n$ is the number of ions involved in the gate, $\delta$ is the symmetric detuning from the center-of-mass (COM) mode, and $\Omega$ is the Rabi frequency. Here we assume that we only couple to the center-of-mass mode of the ion chain and any coupling to other motional modes can be neglected. We also assume that, regardless of the number of ions participating in the gate, the laser power is fixed and uniform across each of the individual addressing beams. Moreover we impose that each of these beams carries the same two frequencies. Taking $\delta=2\eta \Omega$ and a gate time $T = \frac{2\pi}{\delta}$ results is a fully entangling gate among the $n$ ions involved in the operation, where spin and motional degrees of freedom are unentangled at the end of the gate. These parameters for $n=2$ and $n=5$ result in the unitaries of Eqs. (\ref{eq:MS2}) and (\ref{eq:MS5}). \par
We will focus on gates based on Raman transitions (see the following section). In that case, the Rabi frequency is a function of the laser power ($P$) and two-photon detuning ($\Delta$). We consider $\Delta$ to be fixed and not to depend on the number of ions involved in the gate. We then notice that the Rabi frequency, and therefore also the gate time, is independent of $n$. \par It should be noted that there are other ways to implement multi-qubit gates. The implications of the specific choices made here on our results are discussed in section IX.

\section{Error Model} \label{sec:erroromodel}
To accurately compare the performance of two-qubit and five-qubit gates within the toric code, we consider a specific error model with several assumptions. We assume our trapped ion qubits are embedded in the ground state $S$ manifold (either Zeeman or hyperfine), where gates are performed using Raman transitions \cite{scat1,scat2}. \par
There have been many experimental demonstrations of entangling gates based on Raman transitions, and the corresponding sources of error are well classified \cite{fid1,fid2,exp3,exp4,scat3}. Here we choose to disregard all “classical” sources of error that arise from experimental imperfections such as laser phase noise, intensity fluctuations, magnetic field noise, gate-timing errors, among other examples. Thus we consider only the fundamental physical source of error, which for Zeeman and hyperfine qubits is spontaneous scattering of photons from the $P$-level excited states during the gate. \par
The rate of spontaneous scattering during a two-photon Raman transition can be calculated from the Kramers-Heisenberg formula \cite{scat3}. This rate can be shown to be proportional to,
\begin{equation}
    \Gamma \propto \frac{\Omega^2}{P} ,
    \label{eq:scat_rate}
\end{equation}
where $P$ is the laser power at the position of each ion. For each gate then, $p$, the error probability per-ion, is given by a product of this rate with the gate time,
\begin{equation}
    p = \Gamma\tau_{gate} \propto \frac{\Omega^2}{P}\tau_{gate}.
    \label{eq:scat_prob}
\end{equation}
In the gate implementation considered above, the Rabi frequency, gate time, and laser power seen by each ion, are all independent of the number of qubits involved in the gate. Thus Eq. (\ref{eq:scat_prob}) suggests an equal per-ion error probability for two-qubit and five-qubit gates. \par
For every ion involved in a given gate, a scattering event occurs with probability $p$ resulting in a single Pauli error on that ion. An accurate calculation of elastic and inelastic scattering for Zeeman and hyperfine qubits would reveal the respective likelihood of $X$,$Y$, or $Z$ errors \cite{Brown1}. Though for simplicity we will assume here that each occurs with probability $\frac{p}{3}$. Since entangling gates are considerably longer than single-qubit rotations, they are the primary source of error and we choose here to neglect any errors in the latter. \par
Moreover, due to the symmetry between measure-$X$ and measure-$Z$ stabilizers in the toric code, we expect a similar scaling of physical error rates to logical error rates for both $X$ and $Z$ errors \cite{Fowler,Stephens}. Thus we choose to focus only on logical $X$ errors. In terms of error propagation then, we are exclusively concerned with the case where propagation results in any number of $X$ errors. Considering multi-qubit MS gates, we note that this can only occur as the result of a $Z$ error (on any one of the qubits) during a measure-$X$ stabilizer. This is discussed in the following section.

\section{Master Equation Simulation for Multi-Qubit MS Gates} \label{sec:Master Equation Simulation for Multi-Qubit MS Gates}

\begin{figure}
\centering
\begin{subfigure}[h]{\columnwidth}
   \includegraphics[width=.78\linewidth]{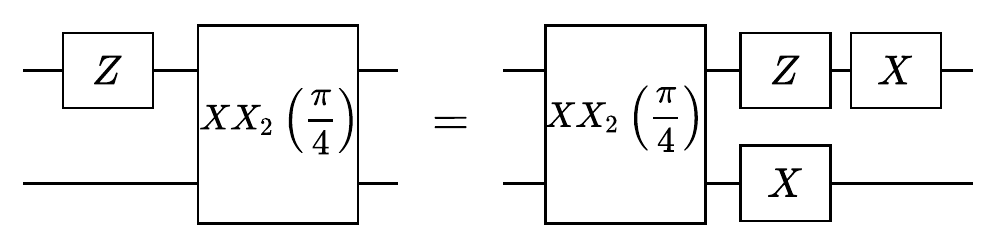}
   \caption{}
   \label{fig:prop2} 
\end{subfigure}

\begin{subfigure}[h]{\columnwidth}
   \includegraphics[width=.78\linewidth]{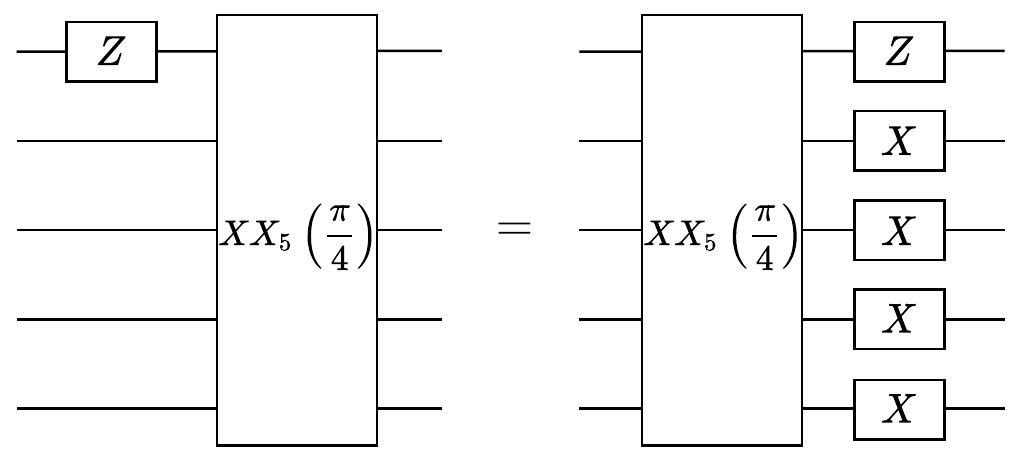}
   \caption{}
   \label{fig:5prop}
\end{subfigure}
\captionsetup{justification=raggedright,singlelinecheck=false}
\caption[PROP]{Circuit depicting propagation of Z error in an XX MS gate for two cases: (a) two-qubit gate (b)
five-qubit gate. As $Y=iXZ$, it's propagation follows similarly.}
\end{figure}

It is important to compare the precise effects of error propagation for both two-qubit and five-qubit gates. For an $XX$-type MS gate, an $X$ error on any qubit commutes through the gate.  $Z$ and $Y$ errors however, do not commute with the gate and propagate as shown in Figures \ref{fig:prop2} and  \ref{fig:5prop}. 

Certain gate errors, such as those due to spontaneous photon scattering considered in this work, may occur at any point during the operation of the gate. In the case of two-qubit gates, when defining an error model for a QEC simulation, it is generally possible to model the error as happening either before or after the gate \cite{Stephens}. With our proposed stabilizer circuits involving five-qubit gates, however, this is not possible as it would not capture the full effects of error propagation. \par

Understanding how errors that occur during the gate propagate can be achieved through a master equation simulation. The Lindblad master equation is given by,  \par

\begin{equation}
\begin{split}
    \dot{\rho}(t) = \frac{-i}{\hbar}[H(t),\rho(t)] + \sum _n \frac{1}{2}[2C_n\rho (t) C_n^\dagger] \\
    -\rho(t)C_n ^\dagger C_n - C_n^\dagger C_n \rho(t)].
\end{split}\label{eq:Lind}
\end{equation}

We use the MS interaction Hamiltonian written in Eq. (\ref{eq:MSH}). As mentioned above, we are specifically interested in learning how $Z$ errors propagate through a MS gate as multiple $X$ errors. Without loss of generality then, we consider a $Z$ error on qubit $\#$1. This error can occur at any point in the duration of the gate, with a fixed error rate. The corresponding collapse operator is $C = \sqrt{\Gamma}Z_1$ where $\Gamma$ is defined in Eq. \eqref{eq:scat_rate}. 

\begin{figure}[h]
\captionsetup{justification=raggedright,singlelinecheck=false}
    \centering
    \includegraphics[width=8.6cm]{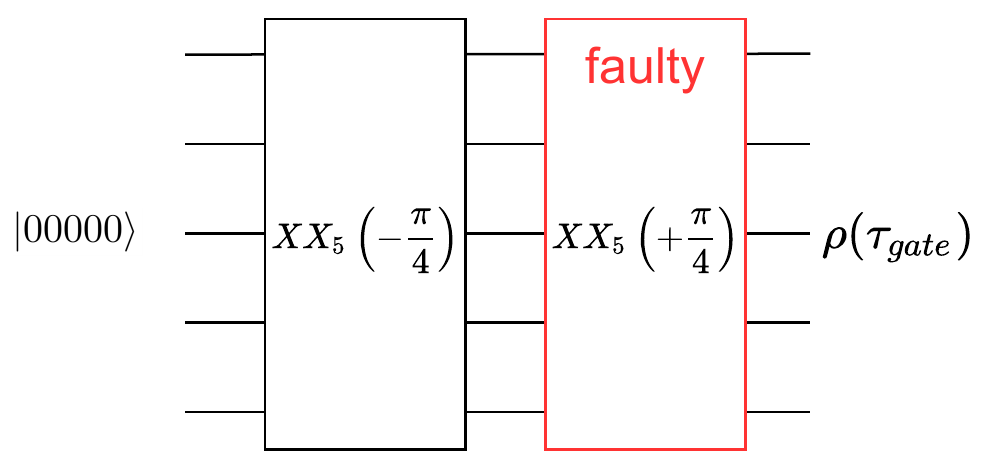}
    \caption[ECHO]{Circuit for determining propagation of Z errors that occur during a five-qubit MS gate. The resulting
    density matrix is calculated using a master equation simulator.}
    \label{fig:lochecho} 
\end{figure}

In order to determine the distribution of propagated $X$ errors, we initialize our state to $\ket{0} ^{\otimes n}$. We run a faulty MS gate (subject to $Z$ error on qubit 1) preceded by an error-free inverse MS gate. This way, if no errors occur in the faulty gate then the $\ket{0} ^{\otimes n}$ state will be recovered. However if a $Z$ error occurred during the faulty gate and propagated as $X$ errors on a given set of qubits, these qubit states will flip to $\ket{1}$. Therefore the diagonal elements of the density matrix, $\rho(\tau_{gate})$, obtained by solving Eq. \eqref{eq:Lind}, reveal the probability distribution of how a single $Z$ error propagated as $X$ errors on each of the qubits. Figure \ref{fig:lochecho} shows this circuit for the case of $n = 5$. \par

We run simulations for two-qubit and five-qubit MS gates, using the software Qutip \cite{Qutip}. We express the results as the likelihood of specific propagated $X$ errors, given that a $Z$ error certainly occurred on qubit 1. The results for the five-qubit case are shown in Table \ref{fig:proptable}. They are organized with respect to two sets of cases. In one set of cases (first two columns), an $X$ error propagated to qubit 1; in the other set of cases (last two columns), no $X$ error propagated to qubit 1. Within each set, there are cases where an $X$ error propagated to two, three, or all four of the other qubits involved in the gate. For each case, all qubit combinations are equally likely; for example, the weight-two error $X_2X_3$ is equally likely as $X_4X_5$ (similarly in the first set, $X_1X_2X_3$ is equally likely as $X_1X_4X_5$).
\begin{table}
    \includegraphics[width=8.6cm]{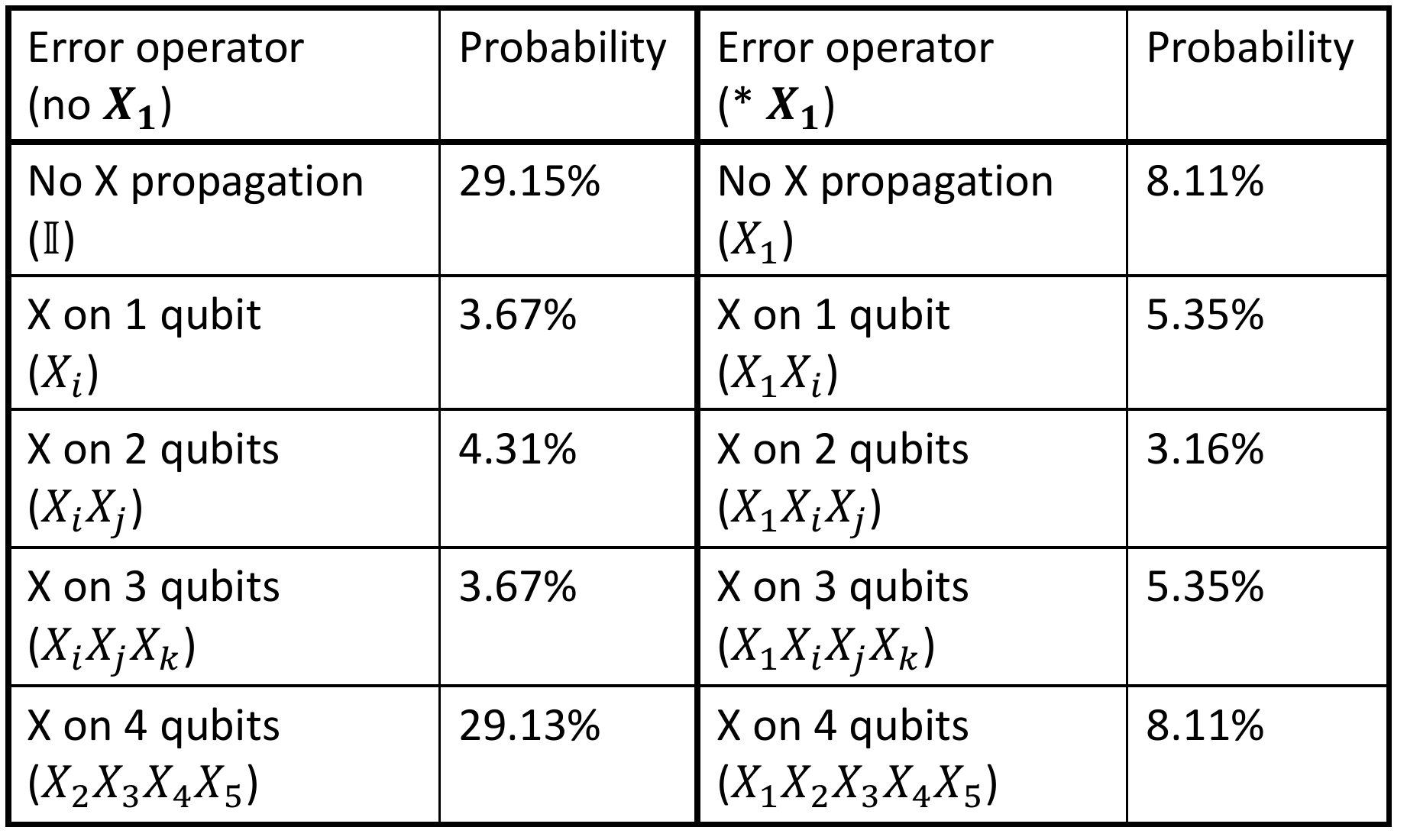}
    \captionsetup{justification=raggedright,singlelinecheck=false}
    \caption[PROPTABLE]{Probability distribution for X error propagation during a five-qubit MS gate. Results obtained by
    simulating a single Z error on qubit $\#$1 occurring with uniform probability over the duration of the gate. The
    first two columns pertain to events where no X error propagated to qubit $\#$1. The last two columns pertain
    to events where an X error did propagate to qubit $\#$1.}
    \label{fig:proptable} 
\end{table}

This distribution of propagated error is used to simulate toric code logical error rates - as described in the following section.

\section{Simulation} \label{sec:simulation}
With a well-defined physical qubit error model, we can simulate the toric code to calculate the FT threshold. To do this, we apply the error model for each entangling operation in the full QEC circuit and account for error propagation. This results in a set of “errors” on a random set of qubits. \par
It is possible that the resulting error syndrome after a single round of QEC is not reliable, due to the presence of “measurement error” on the ancilla qubit. In our error model this occurs via an $X$ error on the ancilla at some point in the circuit - which will flip the measured value of that ancilla. To handle the possibility of an inaccurate syndrome we run $d$ rounds of QEC; this creates a 3-dimensional syndrome that varies in space and time \cite{TC1,TC2}. 

The MWPM algorithm is then left to match points spatially and temporally, where larger distance between two points in any direction makes the pair increasingly unlikely. 

\section{Results} \label{sec:results}
\begin{figure}
\centering
\begin{subfigure}[h]{\columnwidth}
   \includegraphics[width=8.6cm]{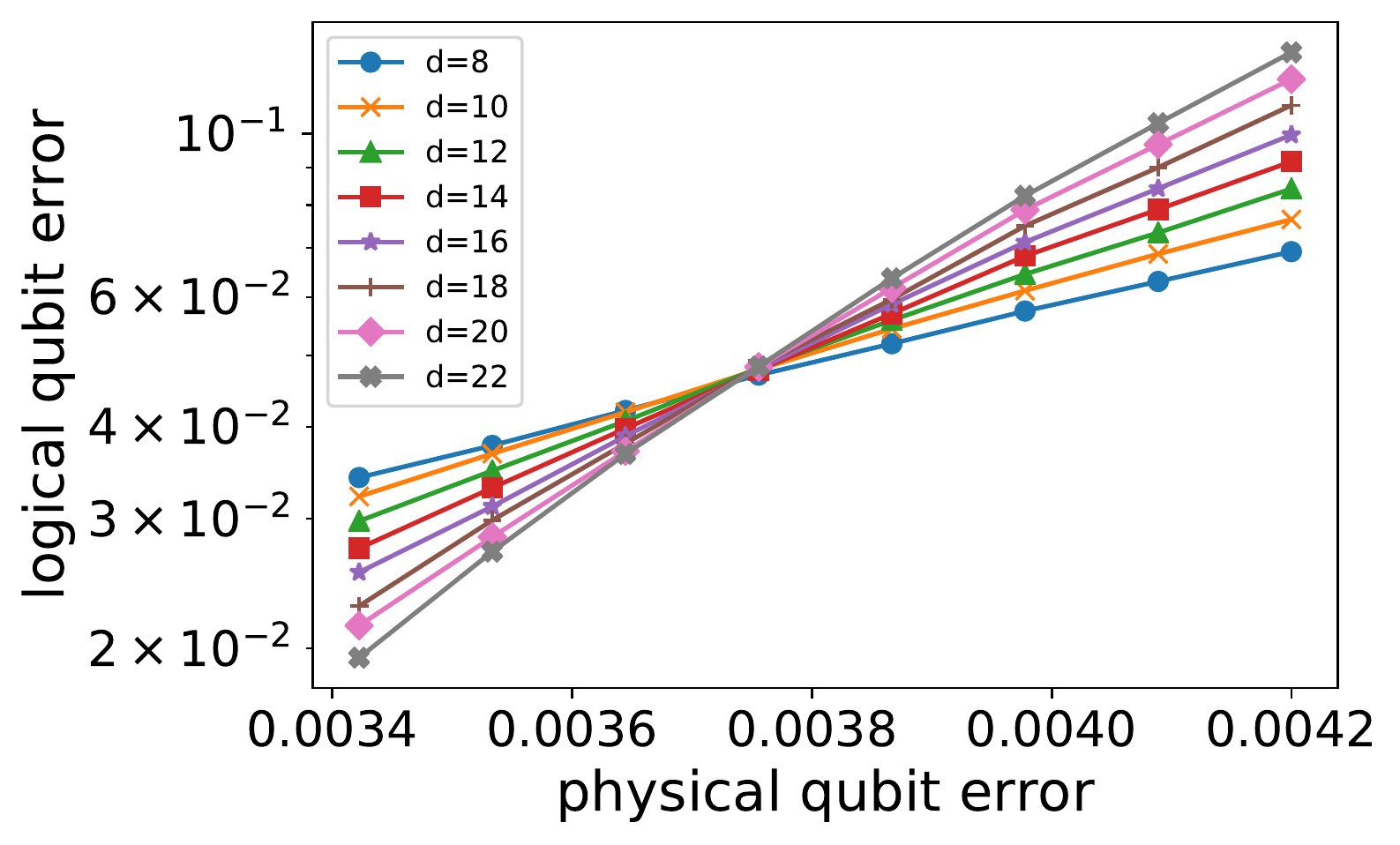}
   \caption{}
   \label{fig:2qubcirc} 
\end{subfigure}
\begin{subfigure}[h]{\columnwidth}
   \includegraphics[width=8.6cm]{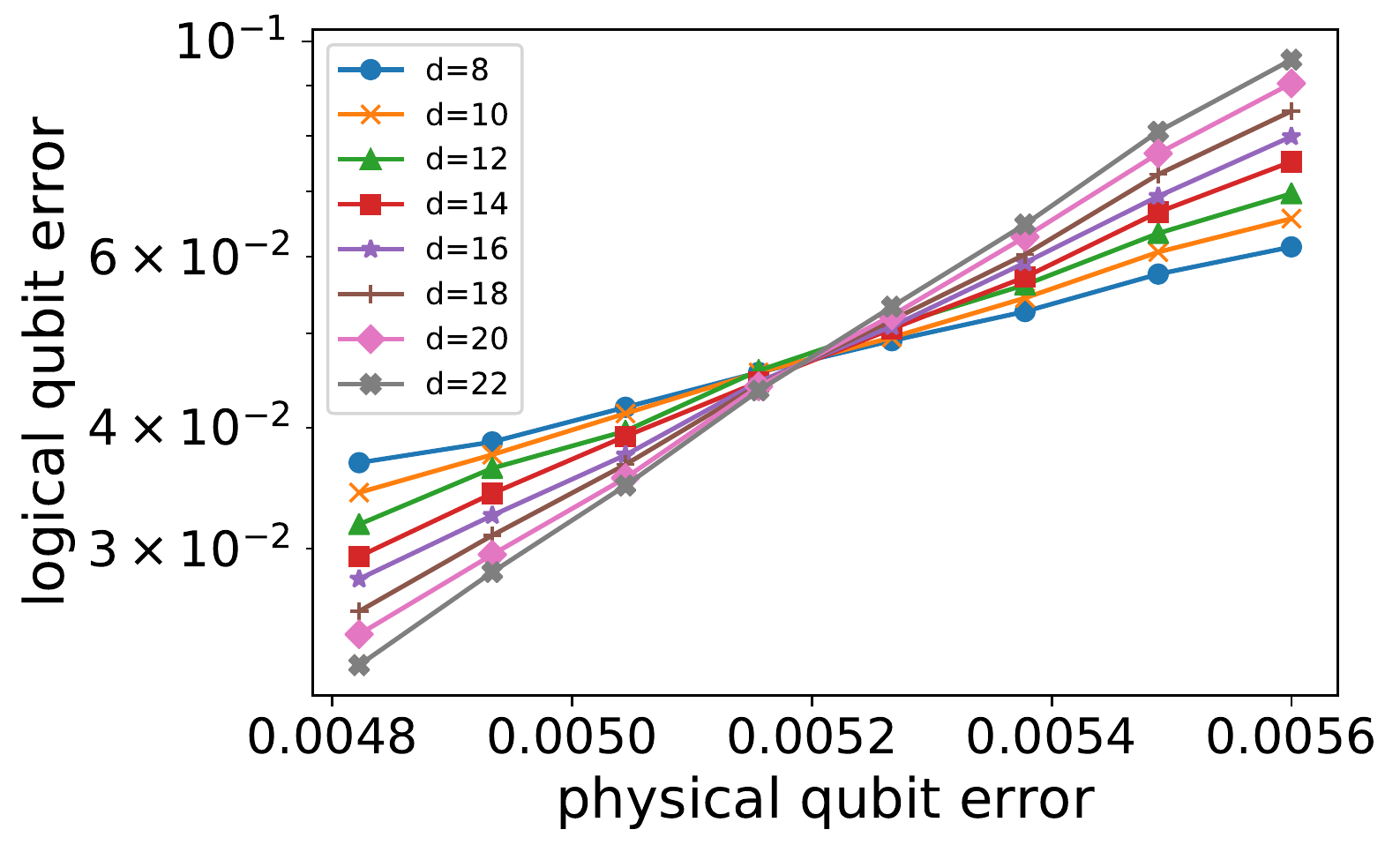}
   \caption{}
   \label{fig:5qubcirc}
\end{subfigure}
\captionsetup{justification=raggedright,singlelinecheck=false}
\caption[RESULTS]{Logical error rate vs. physical error rate after QEC using (a) 2-qubit MS gates (as shown in Figure \ref{fig:4ms}) and (b) 5-qubit MS gates (as shown in Figure \ref{fig:1gms}). The FT threshold is $~.37\%$ in the first case and $~.52\%$ in the second.}
\end{figure}

We plot the performance of the toric code for various distances in Figures \ref{fig:2qubcirc} and \ref{fig:5qubcirc}. The horizontal axis is the physical error probability per qubit, and the vertical axis is the logical error rate of the code. As expected, the logical error rate decreases with the physical error probability. We compare the results for the two-qubit and five-qubit models respectively. The threshold is defined as the physical error probability at the crossing point of the curves. It therefore indicates the upper error bound for when increasing code distance improves logical qubit performance. Each data point is an average of $10^6$ simulation runs, parallelized using the WEXAC computing cluster \cite{WEXAC}. \par 

We observe that the five-qubit gate model results in a threshold of ~$.52\%$. This is an improvement over the two-qubit model, which results in a threshold of ~$.37\%$. We thus note an increase of approximately $40\%$ in the threshold when using the five-qubit gate method. These threshold values are in the range of previous calculations \cite{Wang,Fowler,Stephens,TC1,TC2}, with differences resulting from error model, decoding algorithm, and definition of logical error rate\footnote{It should be noted that these works have expressed the physical error probability as a per-gate value while here it is expressed as a per-qubit value.}. We also note that, aside from the threshold, the five-qubit model gives significantly lower logical error rates. \par
The advantage of five-qubit gates, both in the threshold and logical error rates, can be explained by examining the stabilizer measurement circuits. In the two-qubit model, the ancilla qubit undergoes four sequential gates; by contrast, in the five-qubit model, the ancilla is involved in only one gate. This therefore corresponds to a four times lower probability of ancilla error in the five-qubit case. Moreover, the fact that the ancilla in the five-qubit case is less susceptible to error also results in a lower overall probability of propagated single- and two-qubit errors on data qubits. \par
\begin{figure}
\centering
\includegraphics[width=8.6cm]{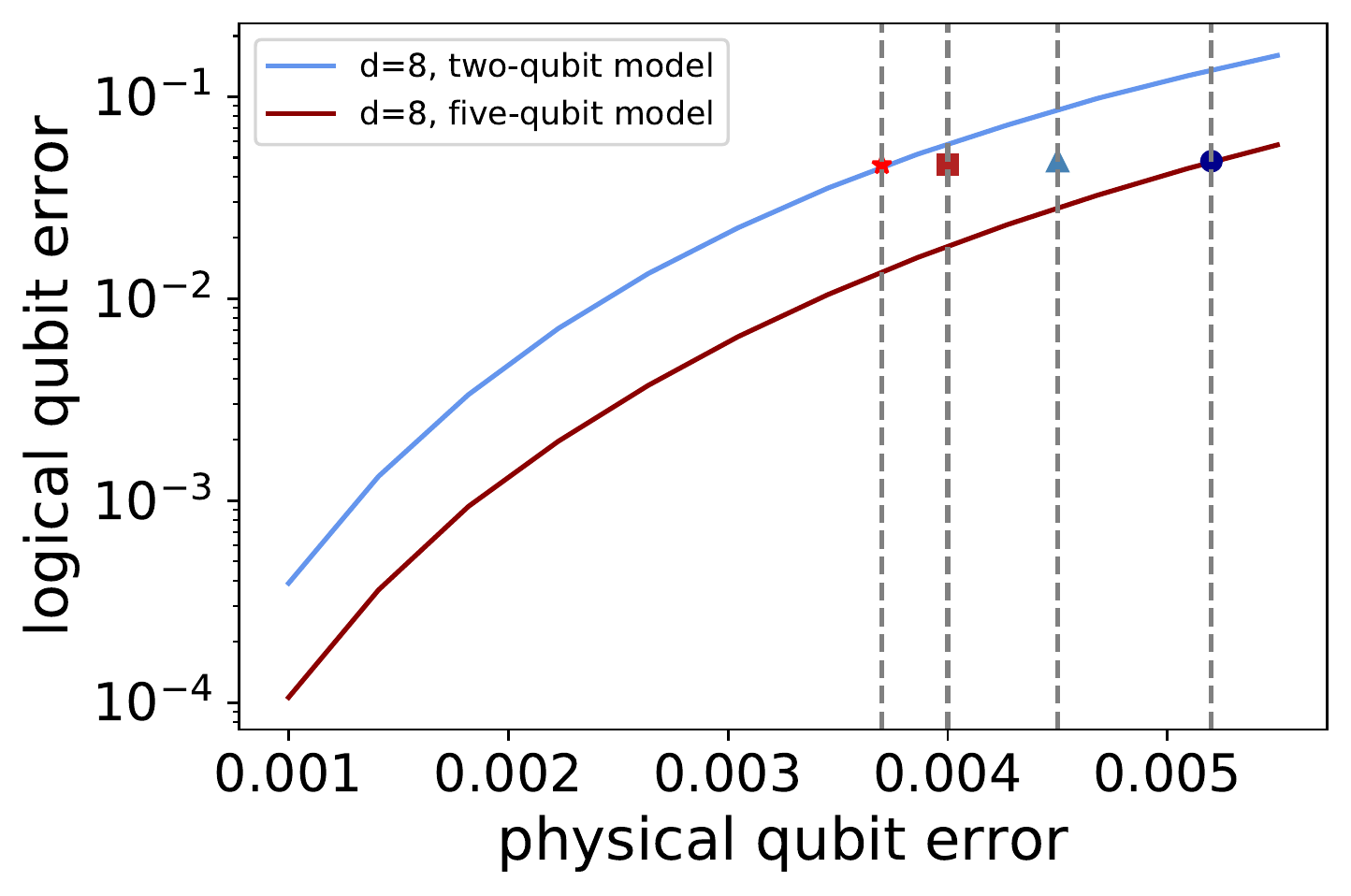}
\captionsetup{justification=raggedright,singlelinecheck=false}
\caption[d8sweep]{A comparison of toric code logical error rates for a distance 8 code using two-qubit and five-qubit gates. The FT threshold for the five-qubit model is at $p=.52\%$ (circle). Parameters of the five-qubit model are incrementally varied to match those of the two-qubit model. These are: propagated single-qubit error probability, propagated two-qubit error probability, and ancilla error probability. The threshold respectively drops to $p=.45\%$ (triangle), $p=.40\%$ (square), and finally to $p=.37\%$ (star).}
\label{fig:d8sweep} 
\end{figure}

In Figure \ref{fig:d8sweep}, we plot the performance of a $d=8$ code for both models at a range of physical qubit error probabilities ($p = .001 \rightarrow .0055$). These parameters reflect a potential near-term implementation of the toric code. The simulation values thus give a practical indication for the performance of the code. At $p=.001$ the five-qubit model gives a ~$3.7$ times lower logical error rate.  \par
We examine here the significance of each error type in the threshold advantage of the five-qubit model. Specifically, we incrementally modify the toric code simulation of the five-qubit model to match the parameters of the two-qubit model. First we impose a probability of propagated single-qubit errors that is equal to its corresponding value in the two-qubit model. We notice that the threshold drops to $p=.45\%$. We then likewise increase the propagated two-qubit error probability, noticing a further threshold drop to $p=.40\%$. Finally we increase the ancilla error probability to match that of the two-qubit model. With all three of these factors included, the threshold reduces to that of the two-qubit model, $p=.37\%$. \par

\section{Assumptions Made} \label{sec:assumptions}
We have shown that using five-qubit MS gates to implement stabilizers in the toric code gives an improvement over the standard two-qubit gate method in both logical error rates and the FT threshold. It is worth stressing that this result is predicated on several assumptions. \par
We have assumed that all entangling operations are mediated by a single mode of the ions' collective motion - specifically the COM mode. We have also assumed that all individual addressing lasers, used to single out ions for entangling gates, have equal intensity and the same spectral content. In other words, the only degree of freedom in implementing an entangling gate is whether to turn each beam on or off. \par
Under these conditions, we have seen that two-qubit and five-qubit gates have equal gate times and spontaneous scattering error rates - and thus equal per-ion error probabilities. We have chosen to consider spontaneous scattering as the only source of error, and neglect other potential sources. \par
While these assumptions reflect current experimental trends, they are certainly not set in stone. For example, recent theoretical proposals \cite{exp4,MQG1,MQG2,MQG3,MQG4} have shown that it may be possible to generalize the standard MS scheme to produce more optimal gates. This may involve modulating the amplitude \cite{MQG1} or frequency \cite{exp4,MQG2} of the laser fields, or using multiple laser frequencies \cite{MQG3}, to couple to many of the ions’ motional modes, not only the COM. In fact, as the size of trapped ion chains increases, it may be increasingly difficult to spectrally isolate a single motional mode - and a multi-mode approach may be the only feasible option \cite{MQG2}. This could, for example, affect the scaling of gate times with the number of ions involved in the gate. \par
One could also consider an experimental apparatus where the total available laser power can be dynamically allocated to only those ions participating in a given gate. In this case, two qubit gates could be implemented with $\frac{5}{2}$ times more laser power per-ion than five-qubit gates. This could allow running two-qubit gates with a $\frac{5}{2}$ reduction of the gate time. In turn, this would result in a $\sqrt{\frac{5}{2}}$ times lower error probability for two-qubit gates and offset the advantage in the FT threshold that we have observed. \par
By focusing on the fundamental source of error, we have shown that multi-qubit gates offer an inherent advantage in error correction.  However, the toric code simulation can also account for wider error models reflecting different experimental implementations. In this way, it can be used to assess the performance of multi-qubit gates on current hardware. \par
Finally, we have not approached the subject of performing stabilizer measurement operations in parallel. In our analysis this had no effect since we did not include idle time errors in our error model. We could thus consider the case where each gate is performed in series with no consequence. This may be a reasonable approximation for qubits with very high coherence times, however it is certainly not optimal. The concept of parallel entangling operations in a chain of trapped ions has been demonstrated \cite{parallelgates}. However, a full experimental procedure for parallelizing gates in the toric code has not yet been proposed.

\section*{Conclusion} \label{sec:conclusion}
The ability to perform multi-qubit ($n>2$) entangling operations is a  useful component of a QC architecture. Indeed \cite{Maslov2} have explored potential applications of global MS operations in various quantum information processing (QIP) protocols. Multi-qubit gates clearly offer an improvement over their two-qubit counterpart by reducing overhead in circuits, and therefore lowering error rates. Here we have shown, with specific focus on a trapped ion system, that the advantage of multi-qubit gates extends to QEC as well. Using the toric code as a benchmark, we have shown that constructing circuits with five-qubit MS gates gives an approximately $40\%$ improvement in the FT threshold over the standard two-qubit approach. \par 

The assumptions made here suit the standard MS scheme; an analysis of how precisely to optimize the gate procedure for QEC codes is a topic for further research.

\section*{Acknowledgements} \label{sec:acknowledgements}
    We thank Kenneth Brown for a useful discussion. This work was performed with the support of the Israeli-Science Foundation. The Willner Family Leadership Institute for the Weizmann Institute of Science, The
Crown Photonics Center and the
Rosa and Emilio Segre Research Award.

\appendix
\section{Stabilizer Measurement Circuit with a Five-Qubit MS Gate} \label{sec:appendix1}
It initially appears that the five-qubit all-to-all entangling gate in Figure \ref{fig:1gms} would create unwanted two-qubit coupling among data qubits. However, in composing this circuit we use the general result of \cite{LW} to write:
\begin{equation}
    \frac{1}{\sqrt{2}}(1+i^{N+E}X^{\otimes N}) = 
    \begin{cases}
    XX_N(\frac{\pi}{4}) & \text{$N$ even} \\
    XX_N(\frac{\pi}{4})R_X(\frac{\pi}{4})^{\otimes N} & \text{$N$ odd}
    \end{cases}
\end{equation}
(Equality is up to a global phase). Where $E = 1$ for even $N$ and $E = 2$ for odd $N$. Thus we see that in the
case of $N = 5$:
\begin{equation}
    XX_5\left(\frac{\pi}{4}\right)R_X\left(\frac{\pi}{4}\right)^{\otimes 5} = \frac{1}{\sqrt{2}}(1-iX^{\otimes N}) 
\end{equation}
From the form of this operator we see that, in conjunction with the single-qubit rotations, the individual two-qubit couplings among data qubits cancel out. From here we can compose the parity-check circuit, knowing that the ancilla qubit starts out in the state $\ket{0}$. Also the four data qubits are in either a $+1$ or $-1$ eigenstate of the stabilizer operator $S_X = X^{\otimes 4}$. Thus we will call the overall code state $\ket{s}$ where $s=\{-1,1\}$ is the eigenstate of this stabilizer operator. Originally the five qubits are in the state: $\ket{0}\ket{s}$ where the first ket denotes the ancilla and the second ket denotes the four data qubits. \par
We apply the operator:
\begin{equation}
\begin{split}
    XX_5\left(\frac{\pi}{4}\right)&R_X\left(\frac{\pi}{4}\right)^{\otimes 5}\ket{0}\ket{s} = \frac{1}{\sqrt{2}}(1-iX^{\otimes N})\ket{0}\ket{s} \\
    &= \frac{1}{\sqrt{2}}(\ket{0}\ket{s}-i\ket{1}S_x\ket{s}) \\
    &=\frac{1}{\sqrt{2}}(\ket{0}\ket{s}-i\ket{1} s \ket{s})
\end{split}
\end{equation}
We then apply a Hadamard gate to the ancilla qubit, leading to:
\begin{equation}
    \frac{1}{2}[(\ket{0} + \ket{1})\ket{s} - i(\ket{0}-\ket{1})s\ket{s}]
\end{equation}
After a final $RX$ gate we have:
\begin{equation}
\begin{split}
    \frac{1}{2\sqrt{2}}[((\ket{0} - i\ket{1}) &+ (\ket{1} - i\ket{0}))\ket{s} \\
    - i((\ket{0}-i\ket{1}) &- (\ket{1} - i\ket{0}))s\ket{s}]\\
    =(\frac{1+s}{2}\ket{0} &+ \frac{1-s}{2}\ket{1})\ket{s}
\end{split}
\end{equation}
(Where the last equality is up to a global phase). We then measure the state of the ancilla qubit which precisely indicates the parity. For even parity ($s = +1$) we will measure the ancilla in $\ket{0}$, for odd parity ($s = -1$) we will measure the ancilla in $\ket{1}$.

\end{document}